\documentclass{article}

\PassOptionsToPackage{super,comma,sort&compress}{natbib} 
\usepackage[preprint]{style} 

\usepackage[utf8]{inputenc} 
\usepackage[T1]{fontenc}    
\usepackage{hyperref}       
\usepackage{url}            
\usepackage{booktabs}       
\usepackage{amsfonts}       
\usepackage{nicefrac}       
\usepackage{microtype}      
\usepackage{amsmath,amsthm,amsfonts,amssymb,amscd}
\usepackage{mathtools}
\usepackage{lastpage}
\usepackage{enumerate}
\usepackage{fancyhdr}
\usepackage{mathrsfs}
\usepackage[x11names]{xcolor}
\usepackage{graphicx}
\usepackage{listings}
\usepackage{subcaption}
\usepackage[english]{babel}
\usepackage{apxproof}
\usepackage{thmtools}
\usepackage{thm-restate}
\usepackage{enumitem}
\usepackage{float}
\usepackage{bm}
\usepackage{wrapfig}
\usepackage{algorithm}
\usepackage{algpseudocode}

\usepackage{sidecap}
\usepackage{siunitx}

\usepackage[title]{appendix}
\usepackage{outlines}

\theoremstyle{definition}

\definecolor{myGreen}{rgb}{0,0.5,0}

\title{The Unbearable Slowness of Being: Why do we live at 10 bits/s?}

\author{
  Jieyu Zheng and Markus Meister \\
  Division of Biology and Biological Engineering\\
  California Institute of Technology\\
  \texttt{\{jzzheng,meister\}@caltech.edu}\\
}

\begin{document}

\maketitle

\section*{In Brief}
Zheng and Meister write about the paradoxical slowness of human behavior. While our senses gather data at $10^9$ bits/s, our overall information throughput is only $10$ bits/s. This stark contrast touches on many fundamental aspects of brain function.

\section{Abstract}
This article is about the neural conundrum behind the slowness of human behavior. The information throughput of a human being is about $10$ bits/s. In comparison, our sensory systems gather data at $\sim 10^9$ bits/s. The stark contrast between these numbers remains unexplained and touches on fundamental aspects of brain function: What neural substrate sets this speed limit on the pace of our existence? Why does the brain need billions of neurons to process 10 bits/s? Why can we only think about one thing at a time? The brain seems to operate in two distinct modes: the ``outer'' brain handles fast high-dimensional sensory and motor signals, whereas the ``inner'' brain processes the reduced few bits needed to control behavior. Plausible explanations exist for the large neuron numbers in the outer brain, but not for the inner brain, and we propose new research directions to remedy this.


\section{Introduction}
``Quick, think of a thing... Now I'll guess that thing by asking you yes/no questions.'' The game `Twenty Questions' has been popular for centuries~\citep{walsorth_twenty_1882} as a thinking challenge. If the questions are properly designed, each will reveal 1 bit of information about the mystery thing. If the guesser wins routinely, this suggests that the thinker can access about $2^{20} \approx 1$ million possible items in the few seconds allotted. So the speed of thinking -- with no constraints imposed -- corresponds to 20 bits of information over a few seconds: a rate of 10 bits per second or less. 

More generally, the information throughput of human behavior is about 10 bits/s.
We review measurements spanning the better part of a century that involve all aspects of human cognition: perception, action, or -- as in the example above -- imagination. 
The general approach is to assess the range of possible actions that a person may execute in a given time. Along the way, one needs a clear criterion to distinguish the action from its noisy variations. This distinction of signal and noise is quantified by Shannon's entropy and ultimately leads to an information rate, expressed in bits/s (see Box).

This information-theoretic approach allows us to compare the speed of processing across different mental tasks and processes, between different neural structures in the same brain, across different species, and between brains and machines. Obviously, this is just one framework by which to characterize human experience, but it offers valuable insights through comparative analysis. 

In particular, our peripheral nervous system is capable of absorbing information from the environment at much higher rates, on the order of gigabits/s. This defines a paradox: The vast gulf between the tiny information throughput of human behavior, and the huge information inputs on which the behavior is based. This enormous ratio -- about 100,000,000 -- remains largely unexplained.

We show that this paradox has implications for many areas of brain science, from neurotheory to assistive technology. A key part of the conundrum is: Why can we only do one thing at a time, whereas our peripheral nervous system processes thousands of items in parallel? We survey various candidate explanations, spanning from evolution to neural circuitry, and find that many of these lead to open questions. We conclude by making suggestions for future neuroscience research to address the conundrum posed by the slowness of being.

\section{The information rate of human behavior}

Take for example a human typist working from a hand-written manuscript. An advanced typist produces 120 words per minute \cite{dhakal_observations_2018}. If each word is taken as 5 characters, this typing speed corresponds to 10 keystrokes a second. How many bits of information does that represent? One is tempted to count the keys on the keyboard and take the logarithm to get the entropy per character, but that is a huge overestimate. Imagine that after reading the start of this paragraph you are asked what will be the next let...

English contains orderly internal structures that make the character stream highly predictable. In fact, the entropy of English is only $\sim$ 1 bit per character \cite{shannon_prediction_1951}. Expert typists rely on all this redundancy: If forced to type a random character sequence,  their speed drops precipitously. So we conclude that the typist operates at

\begin{equation}
    I = 2 \frac{\rm{words}}{\rm{s}} \cdot 5 \frac{\rm{characters}}{\rm{word}} \cdot 1 \frac{\rm{bit}}{\rm{character}} = 10 \frac{\rm{bits}}{\rm{s}}
\end{equation}

Note that we ignored any small variations in the way the keys are struck, such as the amount of force or the exact duration of the keystroke. These differences in the output are considered noise: they do not matter to the task at hand and were likely not planned that way. Thus, they do not constitute different actions and do not contribute to the information throughput. 
We also ignored the typist's occasional yawns and eyeblinks. These can be considered part of the behavioral output, but they are somewhat predictable and contribute at best a tiny fraction of the person's information throughput.

One might argue that the typist is hampered by the speed of the fingers. Perhaps the information rate would be greater when using a motor system that evolved for communication, like the human voice. For the audience to comfortably follow an instructional presentation in English, the recommended narration rate is 160 words per minute~\citep{williams_guidelines_1998}. That is indeed slightly higher than the 120 words/min typing rate, yielding an information rate of 13 bits/s. 

Can one eliminate the need for movements? What would be the speed of pure perception? Some insights come from the competitive sport of "Blindfolded Speedcubing". The competitors solve a $3\times3$ Rubik's cube, but the performance is divided into two parts: the player first examines the cube for a few seconds, and then solves it while blindfolded. Thus the act of perception is separated from the readout phase that requires movement.
(Strictly speaking the perception phase also involves turning the cube over a few times and moving the eyes, but this proceeds at a much more leisurely pace than during the readout phase.)
Both processes are timed, and a recent world record is 12.78 s~\citep{guinnessworldrecordslimited_fastest_2023}: Approximately 5.5 s was used for inspection and the remainder for solution~\citep{tommy_cherry_rubiks_2023}. Because the number of possible permutations of the cube is $4.3 \times 10^{16} \approx 2^{65}$, the information rate during the perception phase was $ \sim 65 \textrm{ bits} / 5.5 \textrm{ s} \approx 11.8 \textrm{ bits/s}$. 
Note also that blindfolded cubers tend to split their total time about equally between perceiving and moving (for another example see~\citep{speedcubing_clips_max_2019}). If the motor system imposed a serious limit on the information throughput, one would expect a very different balance. Instead, it appears that the information rate of perception is nicely matched to the rate of motor output.

A potential concern here is that the cuber has to commit the acquired information into short-term memory, where it has to last for at least 10 seconds, so perhaps the information rate is limited by this need for storage in short-term memory. It is difficult to envision a way to measure perception that does not extend over at least a few seconds of time. 
Instead of shortening that period, we can require a much longer memory interval, and ask whether that alters the rate of perception. 

The answer comes from the realm of memory sports. In the discipline of ``5 Minute Binary'', contestants view pages of binary digits for 5 minutes and then reproduce them from memory within 15 minutes. The current world record holder correctly recalled 1467 digits~\citep{internationalassociationofmemory_5_2024}; the corresponding information rate during memorization is $\sim 5 \textrm{ bits/s}$. A related discipline is ``Speed Cards'': Contestants inspect a random deck of 52 cards for a time of their choosing and have to recall the exact order. However, the recall phase begins only 5 minutes after the deck first appeared, forcing a five-minute memory period. The world record holder in 2018 accomplished the task after 12.74 seconds of inspection~\citep{internationalassociationofmemory_speed_2024}. Because there are $52! \approx 2^{226}$ permutations of a card deck, the information rate during inspection is $\sim 18 \textrm{ bits/s}$. In conclusion, even when the demands on working memory are dramatically increased (from 10 s to 300 s), the rate at which humans use information from the environment remains the same, within a factor of two of our ballpark number of 10 bits/s.

In Appendix \ref{sec:sec:app-behav-more} and Table \ref{tab:table_human} we review additional measurements of human behavioral throughput, ranging from classic laboratory experiments to recent E-Sports competitions. These studies, spanning the better part of a century and divergent fields, all point to a remarkably concordant result: Humans operate at an information throughput rate of about 10 bits/s, and the rates of sensory perception and motor output appear evenly matched. As discussed above, all these tasks involve some criteria for a correct solution, which allows one to separate signal from noise. The range covers many types of human behavior, from games to paid professions, but obviously leaves out others (dancing, dreaming,...) for which an information throughput would be hard to define, let alone measure.

\begin{table}[ht!]
    \centering
    \begin{tabular}[\textwidth]{|p{0.3\textwidth}|p{0.15\textwidth}|p{0.15\textwidth}|p{0.1\textwidth}|}
    \hline
    Behavior/Activity & Time scale & Information rate (bits/s) & References \\
    \hline\hline
    Binary digit memorization & 5 min & 4.9 &~\citep{internationalassociationofmemory_5_2024} \\
    \hline
    Blindfolded speedcubing & 12.78 sec & 11.8 &~\citep{guinnessworldrecordslimited_fastest_2023} \\
    \hline 
    Choice-reaction experiment & min & $\sim$ 5 & \cite{hick_rate_1952, hyman_stimulus_1953, klemmer_rate_1969} \\
    \hline
    Listening comprehension (English) & min - hr & $\sim$13 & \cite{williams_guidelines_1998} \\
    \hline 
    Object recognition & 1/2 sec & 30 - 50 & \cite{sziklai_studies_1956} \\
    \hline
    Optimal performance in laboratory motor tasks & $\sim$15 sec & 10 - 12 & \cite{fitts_information_1954, fitts_information_1964}\\
    \hline 
    Reading (English) & min & 28 - 45 & \cite{rayner_eye_1998} \\
    \hline
    Speech in 17 languages & < 1 min & 39 & \cite{coupe_different_2019} \\
    \hline 
    Speed card & 12.74 sec & 17.7 & \cite{internationalassociationofmemory_speed_2024} \\
    \hline
    StarCraft (e-athlete)& min & 10 &~\citep{guinnessworldrecordslimited_most_2023} \\ 
    \hline 
    Tetris (Rank S) & min & $\sim7$&~\citep{tetrio_tetra_2024} \\ 
    \hline 
    Typing (English) & min - hr & $10$ &~\citep{dhakal_observations_2018, shannon_prediction_1951}\\
    \hline
    \end{tabular}
    \caption{The information rate of human behaviors}
    \label{tab:table_human}
\end{table}

How should one interpret a behavioral throughput of 10 bits/s? That number is ridiculously small compared to any information rate we encounter in daily life. For example, we get anxious when the speed of the home WiFi network drops below 100 megabits/s, because that might compromise our enjoyment of Netflix shows. Meanwhile, even if we stay awake during the show, our brain will never extract more than 10 bits/s of that giant bitstream. More relevant to the present arguments, the speed of human behavior is equally dwarfed by the capacity of neural hardware in our brains, as elaborated in the following section.  

\section{The information capacity of the nervous system} \label{information-nervous}

From the moment that Claude Shannon founded information theory~\citep{shannon_mathematical_1948}, neurophysiologists have applied that framework to signaling in the nervous system. For example, one can view a single neuron as a communication channel. In sensory systems, in particular, we are quite sure what the neurons are communicating about, namely the sensory stimulus, which also happens to be under the experimenter's control. To measure how much information the neuron conveys, one proceeds as described above for behavior: (1) Identify the range of outputs of the neuron across all possible stimuli. (2) Determine what part of that range is ``noise'' vs ``signal''; typically this involves repeating the same stimulus many times and measuring the response variation. (3) Use the expressions for entropy and mutual information to derive an information rate (see Appendix~\ref{sec:app-info-neur}).

For example, the photoreceptor cells in the human eye convert dynamic light input to a continuously varying membrane voltage. Using the approach described in Appendix~\ref{sec:app-info-neur}, one finds that one human cone photoreceptor can transmit information at $\approx 270$ bits/s. The population of 6 million cones in one eye has a capacity of about 1.6 gigabits/s. From that enormous bitstream, the brain of the typist sifts out just the 10 bits/s needed to perform the behavioral task correctly. To represent this degree of filtering, we define the dimension-less ``sifting number'':

\begin{equation}
Si = \textrm{Sifting Number} = \frac{\textrm{Sensory information rate}}{\textrm{Behavioral throughput}} \approx \frac{\textrm{1 Gbit/s}}{\textrm{10 bit/s}} = 10^8  
\label{eq:sifting}
\end{equation}

Unlike photoreceptors, most neurons in the central nervous system communicate via action potentials. The message from one neuron to the next depends entirely on the timing of their spikes. Again, one can adapt the methods of information theory and ask how much information gets transmitted that way. The answer is about 2 bits per spike~\citep{borst_information_1999}. This holds across neurons from many different species and over a wide range of firing rates. 

This ballpark number of 2 bits/spike allows us to estimate the information rate at different stages of sensory processing. For example, the output of the eye consists of 1 million axons from retinal ganglion cells~\citep{kandel_low-level_2021}. With particularly strong stimuli one can drive them at an average rate of 50 Hz. So the capacity of the optic nerve is about 100 megabits/s or less. Note this is 10 times smaller than the capacity of cone photoreceptors: The circuits inside the retina already compress the visual signal substantially, by at least a factor of 10. 

The example of retinal ganglion cells suggests that when driven to high firing rates, individual neurons may transmit hundreds of bits per second. In the mammalian cortex, the average firing rate of a neuron is very low, typically a few spikes per second~\citep{lennie_cost_2003}. However, even under those conditions, a \emph{single neuron} will transmit $\sim$10 bits/s -- equivalent to the information throughput of \emph{the entire human being}. 

\section{The paradox of slow behavior}
To reiterate: 
Human behaviors, including motor function, perception, and cognition, operate at a speed limit of 10 bit/s. 
At the same time, single neurons can transmit information at that same rate or faster. 
Furthermore, some portions of our brain, such as the peripheral sensory regions, clearly process information dramatically faster. 

Before approaching the paradox, we should agree on its magnitude, which sets an expectation for the magnitude of the answer. We need to explain a factor of $Si = 10^8$ , namely the ratio between peripheral information processing and the overall throughput of the brain. Numerically speaking, this may be the largest unexplained number in brain science. This order of magnitude sets a filter for hypothetical solutions. Suppose, for example, that a particular hypothesis explains why people can respond two times faster in one task condition than in another.
That might be enough to publish in
a prominent journal, but the tiny effect size doesn't even make a dent in the paradox. 

Next, we should consider whether one can simply deny the paradox. For many people, the claim that human mental activity is rather slow touches a sensitive nerve. Here we consider some of the objections that have been raised.

\subsection{What about photographic memory?}
It is a popular belief that some people can memorize an entire page of text from a short glimpse, and then recall it as though they were reading an internal photograph. Unfortunately, evidence for photographic or eidetic memory is weak at best. If such people existed, they would sweep the medals in worldwide memory contests, like ``Binary Digits''. Instead, the world champions continue to clock in at a measly 10 bits/s.

One scientific report stands out here, about a single human subject who could fuse a random dot stereogram after each monocular image had been presented on different days \cite{stromeyer_detailed_1970}. Each image had $100 \times 100$ pixels and was presented for 2 minutes, which would require a perceptual rate of 100 bits/s, ten times what we propose. Now a ten-fold discrepancy is nothing dramatic, given the million-fold conundrum we pose. Nevertheless, there are good reasons to question this claim. For example, neural signals from the two eyes are merged already in the primary visual cortex. So the subject's brain would have to somehow inject the memorized image into neurons very early in the visual system, or perhaps even store it there. The claim is so unexpected and potentially revolutionary that it should have triggered many follow-up experiments. No supporting reports have appeared, suggesting that those studies did not go as hoped for. 

Another popular story of photographic memory involves an artist who can paint an entire cityscape at building-level detail after flying over the city in a helicopter~\citep{wiltshire_london_2023}. His paintings are remarkably intricate, but they do not get the details of each building correct~\citep{thedailymail_revealed_2008}. Even if they did, that would still amount to less than 10 bits/s rate of acquisition. Suppose that after a 45-minute helicopter ride~\citep{vocativ_autistic_2017}, the artist drew 1,000 identifiable buildings on a grid, each building with one of 1000 possible styles, and all of them were correctly drawn. In this case, he acquired the information for the painting at a rate of $10^4 \textrm{ bits} / 2700\textrm{ s} \approx 4 \textrm{ bits/s}$.  

\subsection{But what about the rich detail everywhere in my visual scene?} \label{inflation}

Many of us feel that the visual scene we experience, even from a glance, contains vivid details everywhere. The image feels sharp and full of color and fine contrast. If all these details enter the brain, then the acquisition rate must be much higher than 10 bits/s. 

However, this is an illusion, called ``subjective inflation'' in the technical jargon~\citep{oregan_solving_1992, knotts_subjective_2019}. People feel that the visual scene is sharp and colorful even far in the periphery because in normal life we can just point our eyes there and see vivid structure. In reality, a few degrees away from the center of gaze our resolution for spatial and color detail drops off drastically, owing in large part to neural circuits of the retina~\citep{banks_peripheral_1991}. You can confirm this while reading this paper: Fix your eye on one letter and ask how many letters on each side you can still recognize~\citep{rayner_eye_1998}. Another popular test is to have the guests at a dinner party close their eyes, and then ask them to recount the scene they just experienced. These tests indicate that beyond our focused attention, our capacity to perceive and retain visual information is severely limited, to the extent of ``inattentional blindness''~\citep{simons_gorillas_1999, cohen_what_2016}. 

\subsection{But what about unconscious processing?} \label{unconscious}
Another objection to a very low rate of perception is that, by definition, it only takes into account things that we can 
consciously experience and access, so that we can talk about them~\citep{graham_what_1869}. 
Surely, this argument goes, the myriad images that flit over our retina every day must carry a great deal of fine-grained information into the brain. Perhaps this information cannot be accessed consciously as photographic memory, but it is still stored in the brain. 

A specific proposal for unconscious processing involves the development of visual circuits. It is claimed that the visual system learns about the statistics of natural scenes from being exposed to millions of those images over time. Through mechanisms of synaptic plasticity~\citep{bi_synaptic_2001}, the structure of neural circuits responds to visual experiences in the course of early development, which alters the subsequent function of those circuits~\citep{katz_synaptic_1996}. How much information does this process extract from the environment? 

The famous stripe-rearing experiments on kittens can serve as an example~\citep{stryker_physiological_1978}. The kittens were raised in an environment painted with vertical black and white stripes. From the kitten's perspective, because it still got to shake its head, the orientation of the stripes was restricted to perhaps $\pm$20 degrees from the vertical. This did lead to some reorganization of brain wiring in the visual cortex. 
Let us suppose that the kitten brain has completely incorporated the change in visual statistics. 
Whereas in the natural world, all orientations are about equally represented across 180 degrees, in the new environment they are restricted to a 40-degree range. The information associated with such a change in probability distributions is $\log_2 (180/40) \approx 2$ bits.

By perfectly learning the new distribution of angles, the kitten acquired only two bits. 
The same argument applies to other aspects of natural image statistics: The distribution of spatial and temporal frequencies differs from the uniform, but the shape of that distribution can be captured by just a few bits~\citep{olshausen_natural_1996}. This is also true of the distribution of color spectra. In each of these cases, processes of activity-dependent development can learn only a few bits over the animal's lifetime.

Why is the information about scene statistics so limited? 
The main reason is that they are translation-invariant: To good approximation 
the frequency spectrum of a natural image is the same in the bottom left as in the top right. Consequently, it is not necessary to learn separate statistics at each point in the visual field or to specify neural circuitry differently at each point. Nature ``knows'' about this translation-invariance and has exploited it in a clever way: by inventing families of neuronal cell types that are cloned across the visual field~\citep{roska_retina_2014,sanes_types_2015}. For example, the human retina contains many millions of neurons, but they come in only $\sim 100$ types~\citep{grunert_cell_2020}. Cells of a given type are distributed evenly across the retina. And so are the synaptic circuits between cells, because synapses are specified by type identity. In this way, modifying a single gene can alter synaptic circuitry in the same way at thousands of different sites within the visual field~\citep{sanes_synaptic_2020}. Modern image recognition systems have copied this invention in the form of convolutional neural networks, where all units within a layer share the same convolution kernel, thus requiring only a few parameters to be learned~\citep{lecun_gradient-based_1998}.

Our nervous system uses this convolutional trick even when adapting in real-time to a change in image statistics. For example, electrical synapses in the retina can be modified by the neuromodulator dopamine~\citep{bloomfield_diverse_2009}. Thus a change in a single variable, namely the diffuse concentration of dopamine, modifies visual processing everywhere in the visual field. These adaptations occur on the scale of minutes, so again, only a small rate of information flow from the environment is needed for these updates. 

\section{Some implications of ``ten bits per second''}

Having failed at debunking the paradox, one is forced to deal with the recognition that human perception, action, and cognition proceed at a glacially slow pace. 
This insight bears on a number of central problems in neuroscience: from learning and memory, to the evolution of behavior, to the design of neuro-prostheses, or artificial intelligence. Here we merely touch on some of these implications, before considering explanations for the paradox in subsequent sections.

\subsection{How much information is stored in the human brain?}

This is a popular question when people compare the brain to computers or consider the needs for the artificial intelligence systems of the future. It is generally thought that memory is stored in the strengths of synaptic connections, and to some extent in the properties of individual neurons. So we can obtain an upper bound on the brain's capacity by counting the number of synapses ($\sim 10^{14}$) and multiplying by the dynamic range of synaptic strength, estimated at 5 bits~\citep{bartol_nanoconnectomic_2015}, resulting in $5 \times 10^{14} {\textrm{ bits}} \approx 50 {\textrm{ Terabytes}}$ required to specify all synapses. One may want to add a few bits per neuron to specify its functional parameters, but given the small number of neurons ($\sim 10^{11}$), that will leave the estimate at the same order of magnitude. This calculation assumes that any combination of connections is possible and leads to a viable brain, so the upper bound of 50 TB is also a vast overestimate. 

One way to tighten the bound is to consider how information gets into the brain. There are two routes: nature and nurture. The former is implemented by the biological mechanisms of early development, largely steered by a dynamic program of gene expression that unfolds approximately the same in every member of the species. The outcome of this process is encoded in the genome. 
The human genome contains $\sim 3 \times 10^{9}$ bases.
Humans differ from each other at only a tiny fraction of these sites: only a small fraction of the bases encode genes, and not all of the genes are expressed in the brain~\citep{hawrylycz_anatomically_2012}. 
Still, let us suppose generously that all bases in the genome independently control brain development, with each base governing a different dimension of brain connectivity. That places an upper bound on the ``nature'' component at $\sim 6 \times 10^{9}$ bits or $\sim$ 0.8 GB. Again, this calculation pretends that every possible genome will lead to a viable brain, and thus represents a vast overestimate.

Nurture contributes to brain development through the information acquired via the senses. As we have seen, humans at their very best can perceive and memorize about 10 bits/s from the environment. So even if a person soaks up information at the perceptual limit of a Speed Card champion, does this 24 hours a day without sleeping, and lives for 100 years, they will have acquired approximately $3 \times 10^{10}$ bits $<4$ GB. 
After including the information from the genome, this still \emph{fits comfortably on the key-chain thumb drive you carry in your pocket.}

Here again, one could raise the objection that subconscious stimuli somehow drive synaptic development of the brain in a way that bypasses perception and is not limited to the 10 bits/s rate. This is conceptually possible, but no convincing proposal has emerged for what that might be. In an earlier section (\ref{unconscious}), we discussed the scientifically documented case of activity-dependent development driven by the correlation structure of natural stimuli and found that this entails at best a paltry few bits of information. 

One conclusion from this comparison is that the capacity for synaptic representations in the brain (50 TB) vastly exceeds the amount that ever needs to be represented (5 GB) by 4 orders of magnitude. Thus there is plenty of room for storage algorithms that are inefficient due to the constraints of biologically plausible learning rules or massive redundancy.  

\subsection{The speed of life for different species}

How can humans get away with just 10 bits/s? The tautological answer here is that cognition at such a low rate is sufficient for survival. More precisely, our ancestors have chosen an ecological niche where the world is slow enough to make survival possible. In fact, the 10 bits/s are needed only in worst-case situations, and most of the time our environment changes at a much more leisurely pace. This contributes to the common perception among teenagers that ``reality is broken''~\citep{mcgonigal_reality_2011}, leading them to seek solace in fast-paced video games (see Appendix~\ref{sec:sec:app-behav-more}). Previous generations are no exceptions -- they instead sought out the thrills of high-speed sports like skiing or mountain biking. It appears that the everyday tasks feel unbearably slow for these thrill-seekers, so pushing themselves to the cognitive throughput limit is a rewarding experience all by itself.

In other ecological niches, for example, those of snails and worms, the world is much slower still. The relevant threats and opportunities change only slowly, and the amount of computation required to sense and respond is even more limited than in our world. Occasionally, niches intersect with disastrous consequences, as when a snail crosses the highway. One may ask whether there are animals operating at much higher information rates. Candidates might be flying insects that perform aerobatics in turbulent air flow, or birds that navigate through tree clutter at high speeds. 

Remarkably there has been very little research in this area. For example, the visual system of flies has served as the proving ground for information-theoretic measurements in neuroscience~\citep{de_ruyter_van_steveninck_rate_1996,bialek_reading_1991,borst_information_1999}. Yet our survey of researchers in this area could find no attempt to measure the information throughput of the whole fly. Some relevant raw data exist. For example, Theobald et al~\citep{theobald_dynamics_2010} measured the torque that a tethered flying fly produces in response to a horizontal rotation of the visual environment. By analyzing the published time series using the methods reviewed by Borst \& Theunissen~\citep{borst_information_1999} we obtained an information rate of 0.62 bits/s. Note this is 100-fold smaller than the information rate about this kind of stimulus obtained from a single neuron ``H1'' in the fly's visual system~\citep{bialek_reading_1991}. Granted, the torque measurements of Theobald et al~\citep{theobald_dynamics_2010} apply to only one dimension of flight control, and a freely moving insect may operate more efficiently. Nonetheless, it seems likely that the paradoxical contrast between low behavioral throughput and high single-neuron information rates exists in insects as well. 

A serious study of the overall behavioral throughput in different species could be illuminating.
It is generally recognized that gathering information has fitness value~\citep{donaldson-matasci_fitness_2010, kussell_phenotypic_2005}. 
Vice versa information processing entails a metabolic load and thus comes at a fitness cost~\citep{laughlin_metabolic_1998,lennie_cost_2003}. 
But how did different lineages manage this trade-off? Does each species operate at a throughput that is just sufficient for survival? How does that depend on their ecological niche? Does the ``speed of being'' actually contribute to speciation into different ecological niches?  

One species that operates at much higher rates is machines. Robots are allowed to play against humans in StarCraft tournaments, using the same sensory and motor interfaces, but only if artificially throttled back to a rate of actions that humans can sustain~\citep{vinyals_starcraft_2017}. It is clear that machines will excel at any task currently performed by humans, simply because their computing power doubles every two years~\citep{mckenzie_moores_2023}.
So the discussion of whether autonomous cars will achieve human-level performance in traffic already seems quaint: roads, bridges, and intersections are all designed for creatures that process at 10 bits/s. When the last human driver finally retires, we can update the infrastructure for machines with cognition at kilobits/s. By that point, humans will be advised to stay out of those ecological niches, just as snails should avoid the highways.  

\subsection{The Musk illusion}

Many people believe that their internal life is much richer than anything they can express in real time through their mouth or otherwise. 
One can view this illusion as a version of subjective inflation (see \ref{inflation}) generalized beyond visual perception: Because we could engage in \emph{any} one of the $2^{10}$ possible actions or thoughts in the next second, it feels as though we could execute them \emph{all} at the same time. In practice, however, they happen sequentially.

For the most part, this is a harmless illusion. However, when paired with the immense fortune of Elon Musk, the belief can lead to real-world consequences. Musk decided to do something about the problem and create a direct interface between his brain and a computer to communicate at his unfettered rate:
``From a long-term existential standpoint, that's, like, the purpose of Neuralink, to create a high-bandwidth interface to the brain such that we can be symbiotic with AI,'' he said. ``Because we have a bandwidth problem. You just can't communicate through your fingers. It's just too slow.''~\citep{info_clips_elon_2018} 

Based on the research reviewed here regarding the rate of human cognition, we predict that Musk's brain will communicate with the computer at about 10 bits/s. Instead of the bundle of Neuralink electrodes, Musk could just use a telephone, whose data rate has been designed to match human language, which in turn is matched to the speed of perception and cognition.

\subsection{Brain-computer interfaces}

A similar argument applies to brain-computer interfaces (BCIs) intended for patients who are impaired in sensory perception or motor control. For example, in certain forms of blindness, the photoreceptors of the retina die, but the retinal ganglion cells and optic nerve fibers remain intact. One approach to restoring vision has been to implant electrode arrays in the eye and stimulate the ganglion cells there directly with signals derived from a video camera~\citep{stingl_electronic_2013, weiland_retinal_2014}. Of course, this requires transferring raw image information into the peripheral visual system, where -- as reviewed above -- the data rates are gigabits per second. While driven by good intentions, this approach has been altogether unsuccessful: After decades of efforts, all the implanted patients remain legally blind~\citep{ayton_update_2020}. The major companies behind this approach have now gone out of business, and their patients are left carrying abandoned hardware in their eyeballs~\citep{strickland_their_2022,bioregiosternmanagementgmbh_retina_2019}. 

At the same time, we know that humans never extract more than about 10 bits/s from the visual scene. So one could instead convey to the user only the important results of visual processing, such as the identity and location of objects and people in the scene. This can be done comfortably using natural language: A computer translates the visual scene to speech in real time and narrates it to the user according to their needs. Such a device was practical already in 2018~\citep{liu_augmented_2018}, and the intervening developments in computer vision and natural language processing are enabling even more powerful augmented-reality apps for blind people. 

On the motor output side, neural prostheses have been developed that aim to restore some mobility to paralyzed patients. To bridge the gap between the brain and muscles, one can implant a 100-electrode array in some region of the motor or pre-motor cortex, record the spike trains of neurons there, and decode from those signals the user's intended movement~\citep{hochberg_reach_2012, andersen_thought_2019}. The result gets translated into commands for a robot or exoskeleton. These systems have demonstrated some interesting capabilities, for example, a recent benchmark-setting BCI can decode intended handwriting at 90 English characters per minute, or 1.5 bits/s~\citep{willett_high-performance_2021}. A speech BCI can decode intended speech up to 62 words per minute, half the rate of the typist \citep{willett_high-performance_2023}. Such devices could greatly benefit those with conditions like amyotrophic lateral sclerosis (ALS) and Parkinson's disease, who struggle with speech production.  

However, the vast majority of paralyzed people are able to hear and speak, and for those patients language offers a much simpler brain-machine interface. The subject can write by simply dictating text to the machine. And she can move her exoskeleton with a few high-level voice commands. If the robot thinks along by predicting the user's most likely requests, this communication will require only a few words (``Siri: sip beer''), leaving most of the language channel open for other uses. The important principle for both sensory and motor BCIs is that one really needs to convey only a few bits per second to and from the brain, and those can generally be carried by interfaces that don't require drilling holes in the user's head~\citep{chaudhary_brain-computer_2016}.  

\section{Constraints on the speed of cognition} \label{constraints_cognition}
Here we consider some explanations that have been put forward to the two core questions: Why is human cognition so slow? And why does it require so many neurons? 

\subsection{Inefficient neural hardware}

A frequently cited argument is that neurons are simply inefficient devices for information processing, so nature needs to employ huge numbers of them to complete even simple operations. In particular, individual neurons -- unlike transistors -- are subject to random biochemical noise, so perhaps one needs to average over many neurons to get a reliable signal. Also, perhaps the brain includes a large amount of redundancy, with populations of essentially identical neurons, to guard against the loss of neurons from aging or injury. Neither of these explanations seems plausible.

Regarding ``noisy neurons'', the analysis reviewed above (section \ref{information-nervous}) directly addresses that point and shows that a single neuron can transmit several bits per spike about its dendritic inputs. Certain irreducible noise sources-- such as thermal fluctuations of ion channels or vesicle fusion events -- may well limit the rate of information transfer, but the precision of single neurons remains remarkable. 
A single optic nerve fiber can reliably signal the arrival of its preferred feature in the visual stimulus with a precision of 1 millisecond~\citep{berry_structure_1997}. Even deep in the primate cortex, a single neuron can distinguish between two visual stimuli as precisely as the entire monkey~\citep{britten_analysis_1992}. What was once considered irreducible noise at the population level may, in fact, be a reliable signal about variables or inputs that the experimenter cannot control \citep{ringach_spontaneous_2009,stringer_spontaneous_2019}, such as an itch on the monkey's head.


On the subject of redundancy, again there is precious little evidence for duplication of essentially identical neurons. For example, in the fruit fly, many neuron types exist in just two copies, one on each side of the brain~\citep{sterne_classification_2021,aso_mushroom_2014}.
One might argue that fruit flies are cheap and short-lived, so nature simply duplicates the entire organism to ensure robustness. However, the principle applies even to large animals like ourselves. In the primate retina, every point in the visual field is covered by just one neuron of each cell type, with very little overlap~\citep{sanes_types_2015,wassle_cone_2009,reese_design_2015}. In the visual cortex, a small stroke can abolish vision in a section of the visual field. Apparently, there is no redundancy built in for one of the most common forms of brain injury.  

\subsection{Serial vs parallel processing} 
The vast gulf between peripheral and central information rates results in large part from the difference between parallel and serial processing. 
For example, the peripheral visual system performs image processing in a massively parallel fashion. The retina produces 1 million output signals, each of which is the result of a local computation on the visual image. 
Subsequently, the primary visual cortex takes over, using a parallel set of 10,000 modules called ``hypercolumns''. Each of these covers about 0.5 mm$^2$ of the cortical surface, contains $\sim$100,000 neurons, and elaborates a complete set of feature descriptors for a small patch of the visual field~\citep{tso_whither_2009,adams_complete_2007}. 

In contrast, central processing appears to be strictly serial: When faced with two tasks in competition, individuals consistently encounter a ``psychological refractory period'' before being able to perform the second task~\citep{broadbent_perception_1958, pashler_dual-task_1994}. Even in tasks that do not require any motor output, such as thinking, we can pursue only one strand at a time.

It helps to imagine how things could be different. 
For example, one of the celebrated achievements of human cognition is the ``cocktail party effect'', namely our ability to extract one stream of conversation from the chaotic auditory mix in the room and to follow that speaker's string of words~\citep{cherry_experiments_1953, bronkhorst_cocktail-party_2015}. But why do we follow just one conversation, rather than tracking all the conversations in parallel? Certainly, hundred pieces of gossip would be more useful than just one! Our peripheral auditory system operates in parallel, for example processing different frequency channels at the same time, or computing various interaural time delays. 
Yet the selection of a single speaker occurs at a fairly early level, before any word segmentation of the auditory stream~\citep{mesgarani_selective_2012,brodbeck_rapid_2018}.
Why does parallel processing have to end when it comes to the interesting content of audition? From this standpoint, our need to focus attention on just one conversation is a deplorable bug, not a feature.

Another useful perspective comes from the world of competitive games. Chess players decide on the next move, at least in part, based on planning. They consider various options, follow that line several moves into the future, and then evaluate the resulting positions. Why can we not evaluate these possible trajectories in parallel, all at the same time? Given the time limit set on the chess match, a player with that ability would gain a powerful advantage. Yet the psychology of chess is quite clear on the fact that even grandmasters contemplate possible moves serially, one at a time \cite{gobet_psychology_2018}.

So a large part of our paradox resolves to this question: Why is cognition restricted to one task at a time, rather than pursuing many strands -- potentially thousands to millions -- all in parallel? 

\subsection{Evolutionary history}

The only living things with a nervous system are the ones that move around. Accordingly, 
it has been argued for some time that the primary purpose of the brain is the control of movement~\citep{anderson_brain_2016}. 
In early metazoan evolution, a simple nervous system likely served to guide its owner toward food or away from predators~\citep{budd_early_2015}. 
Such a primordial ocean creature might have sensory neurons on the surface by which it can detect an odor gradient. A circuit of interneurons combines this information to ultimately drive the muscles that steer and carry the animal in the desired direction. 
In lineages alive today, one can still recognize that the core parts of the brain are organized around an axis for olfactory navigation~\citep{jacobs_chemotaxis_2012, aboitiz_olfaction_2015}. 

If steering through odor gradients is indeed the original purpose of being, one can see why the brain performs only one such task at a time. After all, the agent is in only one place, it can sense the environment only in that place and must make a decision about which movement to execute. There is no need or even opportunity to simultaneously process multiple paths to food because only one such path corresponds to the current reality. Accordingly, the cognitive architecture is designed to handle that one problem, which is local in space and in time. 

Human thinking can be seen as a form of navigation in a space of abstract concepts. It is like executing movements without activating the muscles. For example, the cerebellum -- long recognized as an organ for motor control -- has certain compartments dedicated to cognition~\citep{koziol_consensus_2014,mendoza_motor_2014}. These areas interact with the prefrontal cortex, instead of the motor cortex~\citep{middleton_basal_2000}.

Another pointer to the primacy of spatial navigation comes from the study of memory athletes~\citep{luria_mind_1987}. Faced with the abstract task of storing long sequences of items, these performers typically imagine a navigation narrative -- like a walk through a familiar neighborhood -- and insert the items to be memorized into locations along that route. During the recall phase of the task, they reenter that ``memory palace'' and ``see'' the items along the route~\citep{krokos_virtual_2019}.
Clearly, these elaborate spatial representations serve as an esoteric hack to fit the task of memorizing binary digits into our natural cognitive architecture. 

In this argument, human thought has co-opted the brain architecture that was originally designed for a jellyfish to follow the smell of decaying flesh in the ocean. With that, our cognition has inherited the constraint of performing only one train of thought at a time. Of course, that still leaves the question how that constraint is embodied in neural architecture.

\subsection{Complexity bottleneck}

Psychological treatments of this subject have invoked a number of metaphors for the serial nature of human cognition, such as ``single channel operation''~\citep{welford_single-channel_1967}, ``attentional bottleneck''~\citep{broadbent_perception_1958,zhaoping_peripheral_2023}, and ``limited processing resources''~\citep{norman_data-limited_1975}. These narratives share a common framework: Sensory systems collect high-dimensional signals at a high rate with many pathways operating in parallel. After considerable filtering and reduction, these signals must compete for some central neural resource, where goals, emotions, memory, and sensory data are combined into low-dimensional decisions. This central neural resource can only execute its functions in series, which constrains the low throughput. But these are not explanations unless one specifies the identity of that limiting neural resource, and to date there aren't any convincing proposals.

The types of cognitive functions that have been associated with the attentional bottleneck do not seem to place high demands on neural hardware. We know this from explicit simulations of neural circuits. For example, Wang~\citep{wang_decision_2008} built a realistic circuit model of decision-making that replicates many of the phenomena observed in perceptual psychophysics experiments on macaques. The circuit contains just 2000 integrate-and-fire neurons, and there is no reason to think that this is a lower limit. Hundreds of these decision-making modules would fit into just one square millimeter of cortex.

A much more demanding cognitive task is image recognition. A neural network model called AlexNet was trained to classify about 1 million photographs into 1000 object categories~\citep{krizhevsky_imagenet_2012,hunsberger_spiking_2015,rawat_deep_2017}. That means extracting 10 bits from an image with 1.2 million bits. This network used 650,000 neurons, the equivalent of just a few square millimeters of cortex. Again, it was not designed to minimize the number of units.
It is difficult to see why we should not have many of these networks running in parallel, rather than a single central pathway for image recognition. 

Another useful point of reference: A fruit fly has less than 200,000 neurons~\citep{raji_number_2021}. Somehow that is sufficient for all the complex operations, including visually-controlled aerobatics, olfactory navigation while walking or flying, social communication, mating and aggression, and more. The human prefrontal cortex alone contains enough neuronal hardware to run 5,000 flies. Why can it not implement a few tasks to run simultaneously? What are the `limiting neural resources''?


It seems that we need to elaborate our model of what these more cognitive brain areas are doing. The current understanding is not commensurate with the enormous processing resources available, and we have seen no viable proposal for what would create a neural bottleneck that forces single-strand operation. We elaborate on this dilemma in the last section.

\section{Outer brain vs inner brain}

The discrepancy between peripheral processing and central cognition suggests that the brain operates in two distinct modes: The ``outer brain'' is closely connected to the external world through sensory inputs and motor outputs.  
This is a realm of high dimensionality: many millions of sensory receptors and muscle fibers, and extremely high information rates. 
The ``inner brain'', on the other hand, operates on a dramatically reduced data stream, filtered to the essential few bits that matter for behavior at any one moment. The challenge for the inner brain is to combine the animal's goals with current inputs from the world and previous memories to make decisions and trigger new actions. The information rates are very low, but the processing must remain flexible because context and goals can shift at a moment's notice. 
A number of interesting research questions emerge regarding the relationship between the inner and outer brain. 

First, how can the inner and the outer brain communicate? The slow inner brain listening to the sensory onslaught from the outer brain seems like 
`drinking from the Hoover Dam'.
The rate of water flow through the Hoover Dam is $\sim 10^8$ times the rate of human drinking, the same ratio as between information rates in the outer vs the inner brain (Eqn \ref{eq:sifting}). 
Presumably, the matching occurs in multiple steps, along some graded continuum of information rate. For example, at the output of the retina, the image information has already been reduced by a factor of 10 or more, leaving only image features that are useful in downstream processing~\citep{roska_retina_2014, field_information_2007, wassle_parallel_2004}. 
Given the large sifting number (Eqn \ref{eq:sifting}), there are many log units of visual compression that remain to be understood.

The retina projects directly to the superior colliculus (SC, also called optic tectum), which is thought to act as a further bottleneck. Neurons in the upper layers of the SC have fine-grained receptive fields similar to those in the retina. By contrast, the lower layers of the SC represent pre-motor signals that reflect the animal's behavioral decisions. Examples include the saccadic eye movements of monkeys, reorienting of the head in owls, or approach vs avoidance of a novel object by rodents ~\citep{basso_unraveling_2021, knudsen_instructed_2002, may_mammalian_2006}. Thus it appears that the SC distills a massively parallel and high-dimensional sensory input into a map of suitable targets for action. The circuit mechanisms underlying this sifting are coming into view~\citep{branco_neural_2020, lee_sifting_2020, de_malmazet_collicular_2023}. Recent work suggests that the SC also plays a role in phenomena of attention that do not involve overt movement~\citep{krauzlis_superior_2013}. Perhaps it contributes to a bottom-up ``saliency map''~\citep{itti_saliency-based_2000} that ultimately directs the attentional bottleneck in cognition. 

Second, what are the principles of neural function on the two sides of the interface?
Taking the water flow analogy further: The design of the Hoover Dam turbines relies on rather different engineering principles from the design of a beer bottle. No one would consider making them of the same materials, given the enormous difference in performance. Yet the brain seems to rely on the same materials throughout: neurons, synapses, and glia. 
The cerebral cortex is touted as the substrate for both rapid and parallel sensory processing (e.g. visual cortex) and for slow and serial cognition (e.g. the prefrontal cortex).  
Very similar neocortical circuits seem to be engaged in both modes. Are we missing some principles of brain design that categorically differentiate the functions of the inner and outer brain?

Comparing research reports from the two sides of the inner/outer brain divide can be a challenge, because the practitioners tend to operate on entirely different assumptions. In the sensory regions, tradition holds that ``every spike is sacred'', because each neuron has its own receptive field, and single spikes are known to convey a few bits of information each about the stimulus~\citep{hubel_early_1998, callaway_local_1998, fitzpatrick_seeing_2000}. For example, the retina has one million output fibers, each with its own visual receptive field. These signals are transmitted with little loss to the primary visual cortex, so one can reasonably argue that the dimensionality of neural activity in V1 is 1 million. 

By contrast, researchers working closer to behavioral output, say in the prefrontal cortex or motor cortex, are often happy to boil the activity of millions of neurons down to just two or three ``latent dimensions''~\citep{gallego_neural_2017, durstewitz_reconstructing_2023, jazayeri_interpreting_2021}. The typical finding identifies a low-dimensional manifold on which the neuronal population vector travels, following simple dynamics. We see here that the paradoxical contrast between inner and outer brain information rates translates to an equally stark contrast regarding the presumed dimensionality of neural activity. Are we to believe that the billion neurons in the primary visual cortex elaborate many channels of visual processing in 10,000 parallel modules, while the billion neurons in the prefrontal cortex just deal with a handful of slow variables, like rules and values, associated with the single task currently at hand~\citep{miller_prefrontal_2000}. 
It seems difficult to accept that two regions of the neocortex with overtly similar cell types and architecture (ignoring the minor differences~\citep{hilgetag_natural_2022}) are organized to operate in such radically different ways. 

Alternatively, the empirically observed difference in dimensionality may be an artifact of experimental design. Most of the studies of inner brain phenomena involve experimental tasks of very low complexity, like a mouse making repeated binary choices between the same two stimuli~\citep{burgess_high-yield_2017}, or a monkey moving a lever along a few possible directions~\citep{shenoy_cortical_2013}. Obviously, the resulting neural representations will be no more complex than what the animal is doing, especially if one averages over many trials~\citep{gao_simplicity_2015}. Under those conditions, one finds by necessity that some low-dimensional manifold accounts for much of neuronal dynamics. However, real-life behavior is not that simple. For example, the Speed Card player, shuffling through the card deck, switches between different neural computations every few tenths of a second: Every eye saccade brings a new card image in focus and triggers a bout of image recognition. These episodes are interleaved with storage in short-term memory, and again with an update of the inner narrative or memory palace that can hold the new card, followed by another saccade. In conversations, we switch quickly between listening and talking, sometimes interleaved with a moment of thinking. When driving, we check the windshield, dashboard, rear mirror, and side mirrors, and process the results in entirely different modes, like estimating distance from the road edge vs reading highway signs. The act of making tea requires 45 different brief subtasks~\citep{land_eye_2006}. Thus, we really perform thousands of different ``microtasks'' on any given day, switching between them as fast as we can saccade~\cite{land_roles_1999, hayhoe_visual_2003}. Each of the microtasks depends on real-time feedback, requiring the sifting of sensory inputs and expansion into motor outputs within sub-second timeframes. The flexible configuration and control of all these data streams seems essential to our cognitive functions. Perhaps the associated routing machinery accounts for the billions of neurons in the inner brain? 

On this background, it would be informative to study the inner brain under such naturalistic conditions where it controls a rapid sequence of distinct microtasks~\citep{gallego_neural_2017}.
Without experimental designs tuned to the specialty of the inner brain, we might fail altogether to discover its essential mechanism.  
Imagine a counterfactual history in which our understanding of the primary visual cortex (V1) was purely based on low-dimensional stimuli: ``We trained a macaque monkey to fixate while viewing a large rotating propeller that reversed direction periodically. After recording from thousands of neurons in the monkey's V1, we discovered that most of the cells responded to this stimulus, so we took a dimension-reduction approach. Fortunately, much of the population activity was captured with just two principal components. And in those reduced dimensions the population vector had rotational dynamics. Finally, we decoded the direction of rotation of the propeller from the rotation of the neural trajectory. After three years of peer review the paper appeared in a high-impact journal.'' 

Given what we know today, none of these hypothetical results touch even remotely on what the visual cortex does. It was only through fine-grained high-dimensional stimuli and single-neuron analysis that researchers uncovered the high-resolution structure of receptive fields, the functional anatomy of their arrangement, and the circuit mechanisms in V1 ~\citep{hubel_early_1998, callaway_local_1998, carandini_we_2005, gegenfurtner_cortical_2003, ohki_specificity_2007}. Returning to the prefrontal cortex -- which contains about the same number of neurons as V1~\citep{gabiNoRelativeExpansion2016,dorph-petersenPrimaryVisualCortex2007} -- it is conceivable that we are missing a similar fine-grained organization. Could there be thousands of small modules in the prefrontal cortex, each dedicated to a specific microtask, much as a hypercolumn in V1 specializes for a small patch of the visual field~\citep{mountcastle_columnar_1997}? This is very speculative, but the only way to resolve our ignorance here is to perform new kinds of experiments.

In summary, we have a sense that major discoveries for a global understanding of brain function are waiting to be made by exploring this enormous contrast between the inner and outer brain. We need to reconcile the `high-dimensional microcircuit' view of the outer brain with its ultimately low-rate information products. And \emph{vice versa} one may need to adopt a more high-dimensional view of the inner brain to account for the computations that happen there to organize behavior. 

\newpage

\section*{Box: Measuring the information rate of human behavior} \label{sec:app-info-human}
How rich is your behavior? To be concrete: How much uncertainty do I have about what you will do in the next second? If you are unconscious and immobile, I can predict very well what will happen over the next second and the uncertainty is low. If you are speaking, typing, or playing tennis, I can make a partial prediction for the next second, assigning probabilities to the various possible actions, but obviously, that leaves some uncertainty. A mathematical measure of this uncertainty is the entropy 

\begin{equation}
    H(A) = - \sum_i {p(a_i) \log_2 p(a_i)} 
    \label{eq:entropy}
\end{equation}

where $p(a_i)$ is the probability of the $i^{\rm{th}}$ action, $\log_2$ is the logarithm to base 2, $A$ is the full distribution of all possible actions, and the result has units of bits. So a prescription for measuring the entropy would be:
\begin{itemize}
    \item Measure many 1-s snippets of behavior, including movements of all joints, sounds coming from the mouth, etc. Call such a snippet an ``action'' $a$.
	\item Distinguish different actions $a_i$ and accumulate their probability distribution $p(a_i)$.
	\item Calculate the entropy according to Eqn \ref{eq:entropy}.
\end{itemize}
	
However, that result is sensitive to how finely I measure your actions. As you scratch your toe, should I measure that movement in microns or centimeters? The finer I sample the movements to distinguish different actions $a_i$, the larger the entropy according to Eqn \ref{eq:entropy}. So what is a natural bin size for actions?

This dilemma is resolved if I give you a specific task to do, such as typing from a hand-written manuscript. Now we can distinguish actions that matter for the task from those that don't. For example, two different keystrokes are clearly different actions. But if you strike the key in 91 ms vs 92 ms, that variation does not matter for the task. Most likely you didn't intend to do that, it is irrelevant to your performance, and it is not what I want to capture as ``richness of behavior''. 

So the presence of a concrete task leads to a natural binning of the actions, in this case into keystrokes. It also allows us to separate the \emph{signal}, namely the entropy that determines task performance, from \emph{noise}, namely the part that does not. The difference between total entropy and noise entropy is the mutual information

\begin{equation}
    I = H(A) - H(A|C) 
    \label{eq:mutual-info}
\end{equation}

where $C$ stands for the entire context of the measurement, namely the task you are solving and the entire history of your actions up to the present point in time. $H(A|C)$ is the conditional entropy of your actions in the next second, given that I know the entire context: 

\begin{equation}
    H(A|C) = - \sum_{(a, c) \in A \times C} {p(a, c) \log_2 p(a|c)} 
    \label{eq:cond-entr}
\end{equation}

where $p(a|c)$ is the conditional probability that you perform action $a$ given context $c$, and $A \times C$ the joint set of all possible actions and contexts. 

In many examples discussed in the text, the task performance is considered perfect, for example, a manuscript without typos, or a perfectly solved Rubik's cube. A performance without errors means $H(A|C)=0$ because the actions can be predicted perfectly from knowledge of the context. 

We adopt the use of information theoretic measures here, because they allow a comparison of behavioral performance in neurons, networks, entire organisms, and machines using the same units of bits per second. In simple terms, a behaving agent chooses one among many possible actions given a set of inputs, and the information rate measures how much that range of actions gets constrained by the decisions. Information theory is not the only statistical framework one could employ, and we don't consider it a magic bullet~\citep{shannon_bandwagon_1956}.

In the same vein, many of the information numbers quoted here should be seen as rough estimates. In some cases, the prior or posterior probability distributions are not known precisely, and the calculation is only based on mean and variance. Those estimates are probably reliable to within a factor of 2 either way. Fortunately the main topic of the article involves much larger factors. 

\newpage
\begin{appendices}
    
\section{The information rate of human behavior: more examples} \label{sec:sec:app-behav-more}

In the main text, we review the information rate of a human typist, an example of real-time processing. Younger readers who are unfamiliar with typewriters will find it reassuring that the same number also applies to experienced video game players. 

\subsection{Video games and motor capacity}
The early video games are good case studies, as it is possible to enumerate all possible outcomes based on the discretized graphics and limited types of button presses. In the game Tetris, for example, players are required to stack different shapes of four-unit pieces (called tetrominoes) being dropped at a constant rate, one at a time from the top of the screen. To prevent the pieces from accumulating, players have to arrange the pieces to fill each ten-unit row that will be removed once completed. 
To estimate the upper bound of information, we can assume that every placement of a tetromino -- its orientation and location in the row -- is equally probable. Of all the tetrominoes, the T-shape piece has the highest number of possible orientations (4) $\times$ locations (9) to choose from. Multiply this by the highest human placement rate (3-4 pieces per second on average for the highest tier gamers~\citep{tetrio_tetra_2024}), and you will get an information rate of around 7 bits/s.  

For convenient comparisons across games, we can measure the prowess of a player by actions per minute (APM) to estimate information rates. While the best Tetris players perform at $\sim$200 APM~\citep{tetrio_tetra_2024}, professional StarCraft players reach 1000 APM during intense battle sequences~\citep{guinnessworldrecordslimited_most_2023}. Again there is a considerable amount of redundancy compared to typing English, because only a few important hotkeys are frequently used for the game, and many presses serve the same given order. Thus, the entropy of the keystroke string in the game is very likely to be lower than that of English letter strings (1 bit/character). We will generously credit the StarCraft champion with zero errors and 1 bit/character as the typist, and conclude that the information throughput at the very height of battle is at the same order of magnitude as a typist during a regular work day, at $\sim$ 16.7 bits/s.

Granted, our fingers did not evolve for typing, and one may argue that the human motor system can be measured by tasks with more freedom of movement than button bashing. Fitts~\citep{fitts_information_1954} trained the participants to perform four different types of motor tasks that involved relocating one's forearms rapidly between confined target areas. At their best performance within a short spree of 20 seconds, the well-practiced subjects reached 10-12 bits/s in all four tasks. Different speed or accuracy requirements yielded an invariant human motor capacity at 10 bits/s~\citep{fitts_information_1964}. 

\subsection{Language}
Of the entire human motor system, our vocal tracts are the most specialized in communicating information. Analyses of speech recordings in 17 languages peg human speech at a universal transmission rate of 39 bits/s~\citep{coupe_different_2019}. 
Here, the information rate is the product of information density and speech rate, but the two are inversely correlated because of the speed-accuracy trade-off: fast speech tends to consist of more frequently-used but less informative words~\citep{mackay_problems_1982, cohen_priva_not_2017}. Speakers tend to slow down when they introduce contextually unexpected vocabulary. 
In addition, this number does not take into account much of the redundancy in language: as acknowledged in the methods section of the article, the true information rate is lower. Still, the estimate is in the same order of magnitude as our ballpark number.

Even in cases of minimal movement, the speed of language processing remains in the same range. During reading, for example, one fixation typically lasts 200-250 ms and covers 7-9 letters~\citep{rayner_eye_1998}. This yields an average reading speed of 28 - 45 English characters per second or bits/s. Similar to fast speech, a speed reader can skim through the text and triple their rate up to 700 words per minute or $\sim$ 60 characters/s, but comprehension is compromised 
\citep{masson_conceptual_1983}. If one includes this error rate, the rate of reading remains considerably below 60 bits/s.

\subsection{Perception}
What about innate or quickly learned responses to visual stimuli? The extensive research on human reaction time explored the upper limit of such responses. In a choice-reaction experiment by Hick~\citep{hick_rate_1952}, participants made movements corresponding to the flashes of two to ten light bulbs. The light bulbs flickered in random order; therefore, to press the correct corresponding key was to make one out of 2 - 10 equally probable choices. As the number of lamps increased, the reaction time -- at hundreds of milliseconds -- was found to increase logarithmically. Hyman~\citep{hyman_stimulus_1953} repeated the experiment in units of bits, which resulted in a linear relation and a constant slope -- the information of the choices divided by the reaction time, or the information rate of the reaction. Over thousands of trials, ``the rate of gain of information is, on the average, constant with respect to time'', at 5 bits/s. Different experimental designs reached a similar conclusion~\citep{klemmer_rate_1969}, with information rates ranging from a few to ten bits/s. 

In cases where the need for motor output is eliminated, the rate might be slightly higher, at 30-50 bits/s for object recognition~\citep{sziklai_studies_1956}. In this study, each participant viewed multiple line drawings of objects that were shown simultaneously. When a frame of 16 objects was shown for 500 ms, all 50 participants recognized at least one independent object; the majority identified two, and only one individual managed to name three in a trial. 
Sziklai~\citep{sziklai_studies_1956} then modeled object recognition as selecting from a repertoire of a thousand basic picturable nouns -- based on the vocabulary an adult English speaker should know. Thus each named item corresponds to 20 bits of information. Divide 20 bits by 500 ms and we get 40 bit/s on average.  

Table \ref{tab:table_human} summarizes the human information rates of different tasks discussed in this article.

\section{The information capacity of neurons} \label{sec:app-info-neur}
Here is a back-of-the-envelope calculation of information capacity for neurons that communicate with either continuous potentials or action potentials.

\subsection{Photoreceptors}
The human retina has $\sim6$ million cones and $\sim100$ million rods. We consider only the cones because they account for almost all of everyday vision. The cell's membrane potential has a maximum range of $\sim$100 mV. Under good lighting conditions, the noise in that potential is $\sim$1 mV, in the sense that the output under repeated identical stimulation will fluctuate by that amount. So the photoreceptor can encode about 100 different messages at any given time, which corresponds to an entropy of $\sim$ 7 bits. 
What is the rate per unit time of such messages? Photoreceptors are rather slow neurons, owing to the biochemical amplifier that turns tiny photons into large membrane currents. The bandwidth of the cone signal is $\sim20$ Hz, meaning it can change about 20 times per second~\citep{schneeweis_photovoltage_1999, baudin_s-cone_2019}. Using the Shannon-Hartley theorem for capacity of a Gaussian channel~\citep{cover_elements_2012} one finds that the information capacity of a single cone is $\sim \log_2{100^2} \textrm{ bits} \times 20 \textrm{ Hz} \approx 270 \textrm{ bits/s}$. The entire retina contains $\sim6$ million cones. They all have separate receptive fields and can be stimulated independently. Thus the information capacity of the cone population of one eye is $6 \times 10^6 \times 270 \textrm{ bits/s} \approx 1.6$ gigabits/s.

\subsection{Spiking neurons}
The typical spiking neuron can be viewed as a point-process channel: It converts a continuous time-varying input, like the synaptic current $s(t)$, into a discrete train of spikes at the output, $r=\{t_i\}$. Several approaches have been developed to estimate the resulting information rate~\citep{borst_information_1999}. A popular one is by ``decoding the spike train'': Propose a model to estimate the stimulus $s_{\rm{est}}(t)$ from the spike train; the simplest such model is linear filtering of the spike train with some kernel $k(t)$, 

$$
s_{\rm{est}}(t) = \sum_i k(t-t_i)
$$

From an experiment that measures both $s$ and $r$, optimize the decoding kernel $k(t)$ so as to minimize the difference between $s(t)$ and $s_{\rm{est}}(t)$. Finally measure the mutual information between $s(t)$ and $s_{\rm{est}}(t)$. These are both continuous functions, so the Shannon-Hartley theorem applies, though one needs to be careful along the way in managing lower and upper bounds so as to maintain a conservative estimate.
For many neuron types,  the information capacity of the spike train has been measured at 1-4 bits/spike~\citep{borst_information_1999}. 

\subsection{Capacity vs information rate}
For a given information channel, such as a human being or a single neuron, the information transmitted will depend strongly on the choice of input. For example, if a typist is given a manuscript with endless repeats of the same letter, she cannot convey much information, simply because the source has low entropy. Likewise, if a photoreceptor is driven by light with only small and slow variation, then it will convey little information. The {\emph{capacity}} of a channel describes the maximal information rate achievable using the optimal distribution of the input. For a typist that would mean well-legible handwriting in a language she has practiced, good illumination, high contrast, ergonomic chair, etc. For a photoreceptor that means high illumination, high contrast, and rapid variation at frequencies that span the whole bandwidth.

In normal function, both people and single neurons are rarely driven at the maximal rate specified by the capacity. People reach their capacity under special professional conditions (typist) or the pressure of competition (Starcraft player). Sensory neurons are found to reach their capacity preferentially during biologically important tasks, like the discrimination of mating calls (frogs), or the real-time control of aerobatics (flies)~\citep{bialek_reading_1991, rieke_naturalistic_1995}.
\end{appendices}

\newpage
\section*{Acknowledgments}
For discussions and comments on the early stage of the manuscript, we thank Frederick Eberhardt, Michelle Effros, Florian Engert, Mehrdad Jazayeri, Christof Koch, Ueli Rutishauser, and Anthony Zador. We also thank alphaXiv for providing a platform for open discussions on our preprint, where the comments received were intriguing and helpful for the revision. 

\section*{Funding}
MM was supported by grants from the Simons Collaboration on the Global Brain (543015) and NIH (R01 NS111477).

\section*{Declaration of Interests}
The authors declare no competing interests.

\section*{References}
\renewcommand{\bibsection}{}\vspace{0em}
\bibliographystyle{vancouver}
\bibliography{references} 

\end{document}